\documentstyle[12pt]{article}
\begin{document}
\renewcommand{\thefootnote}{\fnsymbol{footnote}}
\setcounter{page}{1}
\begin{titlepage}
\begin{flushright}
\large
INFN-FE 08-94 \\
July 1994
\end{flushright}
\vspace{3.cm}
\begin{center}
{\Large\bf On Instanton Contributions to the $\tau$ Hadronic Width}\\
\vspace{0.5cm}
{\large
Vakhtang Kartvelishvili${}^{(a)}$\footnote
{E-mail: vato@vxcern.cern.ch, vato@kheta.ge } and
Murman Margvelashvili${}^{(a,b)}$\footnote
{E-mail: mm@infn.ferrara.it, mmm@kheta.ge }}

\vspace{0.4cm}

{\em ${}^{(a)}$ High Energy Physics Institute, Tbilisi State University
380086, Tbilisi, Georgia

\vspace{0.2cm}

${}^{(b)}$ INFN sezione di Ferrara, I-44100 Ferrara, Italy
 }

\vspace{0.3cm}

\end{center}
\vspace{1.0cm}
\begin{abstract}
Comparing the results of finite energy and Borel transformed sum rules
for the difference of vector and axial vector current correlators
we evaluate the instanton contributions to the $\tau$ hadronic width.
In contrast to an explicit theoretical calculation we find that the
instanton contributions to the $\tau$ hadronic width are much smaller
than the standard nonperturbative corrections.
\end{abstract}
\end{titlepage}
\renewcommand{\thefootnote}{\arabic{footnote}}
\setcounter{footnote}{0}
\vspace{0.4cm}
\newpage

{\bf I.} \,\,Since it has been realised that an accurate theoretical
calculation of
$\tau$ hadronic width is possible, considerable work has been done for
improving the accuracy of experimental and theoretical determination of
this observable. The main aim of these studies is to determine the QCD
coupling $\alpha_s$ at the level of accuracy competitive with that at the
$Z^0$ pole. The precision of $\alpha_s$ measurement is mainly limited by
the ignorance of $O(\alpha_s^4)$ perturbative corrections. There are,
however,
nonperturbative contributions which should also be taken into account
\cite{bnp,dib,p,al}.

The nonperturbative effects parametrised by QCD condensates \cite{svz}
are estimated by means of the operator product expasion (OPE) in QCD
vacuum. They were shown to be on the level of $O(1\%)$ of $\tau$ hadronic
width [1-4].

Recently contributions due to instantons have also estimated \cite{np,bbb}.
While in \cite{np} they were found to be negligible, ref.\cite{bbb}
resulted in quite sizeable instanton correction.
This calculation, in spite of numerical uncertainties, leads to two
disappointing consequences. First, it implies that nonperturbative
contributions cannot be reliably calculated and thus puts limit to the
accuracy of $\alpha_s$ determination. Next, the strong instanton
contributions signal the breakup of the OPE series and can lead to
considerable inconsistencies in QCD sum rules
(SR) approach making the finite energy (FESR) \cite{fesr} results
incompatible with that of Borel
transformed (BTSR) sum rules \cite{svz}.

In the present paper we exlore the consequences of instanton calculations
\cite{bbb} for the difference of $\tau$ decay widths into an
even and an odd numbers of pions, which is an observable sensitive to the
nonperturbative effects. We exploit the fact that in this difference
the perturbative contributions cancel out while nonperturbative effects are
enhanced. Estimating instanton corrections for this quantity we are
able to impose bounds on corresponding contributions in $\tau$ hadronic
width.To evaluate the instanton contributions we use the observation
\cite{bbb} that  in BTSR they should be strongly (factorially)
suppressed while they are present in FESR. Using the spectral
densities extracted from experimental data on tau decays,
we compare the results obtained from FESR and Borel SR. At
the present level of accuracy we see no sign of instanton conributions
and thus conclude that nonperturbative effects in $\tau$ hadronic width
are mainly due to  dimension 6 and 8 condensates.

{\bf II.} \,\,Consider the difference of $\tau$ decay widths into even and
odd numbers of pions, normalized as usual to the leptonic decay width:
\begin{equation}\label{rdef}
R_{\tau,V-A}=\sum_n{\frac{\Gamma(\tau\to\nu_{\tau}+2n\pi)-
\Gamma(\tau\to\nu_{\tau}+(2n+1)\pi)}{\Gamma(\tau\to\nu_{\tau}+e\nu_e)}}
\end {equation}

In the following we will be working in the chiral limit, which was also
used for instanton calculations \cite{bbb}. Then the considered quantity
can be expressed as
\begin{equation}\label{r}
R_{\tau,V-A}=12\pi |V_{ud}|^2\int_0^{M_\tau^2}{\frac{ds}{M_\tau^2}}
{(1-\frac{s}{M_\tau^2})^2}{(1+\frac{2s}{M_\tau^2})}{\rm Im}
\Pi_{V-A}(s+i\epsilon)
\end{equation}

Here $\Pi_{V-A}(s)$ is the difference of the vector and axial vector
current correlators, defined as
\begin{equation}\label{pimn}
(-g^{\mu\nu}q^2 + q^\mu q^\nu) \Pi_{V-A}(q^2) =
 i\int d^4x\,e^{-iqx}<0|T[V^\mu(x)V^\nu(0)-
   A^\mu(x)A^\nu(0)]|0>
\end {equation}
with $V^\mu=\bar u \gamma^\mu d$ and $A^\mu=\bar u \gamma^\mu \gamma_5 d$,
while ${1\over{\pi}}{\rm Im}\Pi_{V-A}(s)=\rho(s)$ is the
difference of even- and odd-pion spectral densities, measured in $\tau$
decays. In the following for the CKM matrix element we assume $|V_{ud}|=1$

Some time ago we have used the experimental data obtained by ARGUS
collaboration for extracting the spectral function $\rho (s)$ within
certain quite
general assumptions \cite{web,wef}. We have used explicitly linear fitting
procedure, which allows to trace easily the propagation of experimental
errors from differential invariant mass distributions to the weighted
integrals over $\rho(s)$. Substituting this spectral function
into (\ref{r}) we can estimate $R_{\tau,V-A}$. Let us use the
notation of ref. \cite{bnp}
\begin{equation}\label{rdel}
R_{\tau,V/A}=(3/2)(1+\delta^0+
\delta^6_{V/A}+\delta^8_{V/A}),
\end {equation}
where $\delta^0$ stands for the perturbative correction and $\delta^6,
\delta^8$ represent dimension 6,\,8 condensate contributions; Taking the
difference ofdecay widths into vector and axial-vector decay channels,
 we obtain
\begin{equation}\label{dour}
R_{\tau,V-A}={\frac{3}{2}}(\delta^6_{V-A}+\delta^8_{V-A})=0.07 \pm 0.05
\end {equation}
This value can be compared to the difference of branching fractions
(\ref{rdef}) obtained using recent ALEPH data \cite{al}:
\begin{equation}\label{alexp}
R_{\tau,V-A}=0.02 \pm 0.09
\end {equation}
The error quoted in (\ref{alexp}) is obtained by summing up the errors
in partial widths in quadrature. Thus we have ignored (obviously strong)
correlations among separate branching fractions, but we hope that the error
in (\ref{alexp}) is right at least by an order of magnitude.

In \cite{bbb} the instanton contributions to $R_{\tau,V/A}$
have been evaluated by calculating the exponential correction to the
coefficient function of the six-quark operator in the OPE of $\Pi_{V,A}$.
Unlike ref.\cite{np}, the authors find sizable instanton
contributions to the full hadronic decay width of $\tau$. These contributions
are enhanced even more in the difference $R_{\tau,V-A}$
and from ref.\cite{bbb} one can find:

\begin{equation}\label{dbbb}
\delta_{V}^{inst}-\delta_{A}^{inst}=0.06-0.1\,\,\Rightarrow \,\,
R_{\tau,V-A}^{inst}=0.09\div0.15
\end {equation}

Comparison of this value with the result of the integration of experimental
spectra (\ref{dour}) and/or the measured branching fractions (\ref{alexp})
leads to the conclusion that $R_{\tau,V-A}$ is completely dominated by
the instanton contributions leaving no space for standard condensate
terms.
However, in view of numerical uncertainties in instanton calculations this
result should be tested more carefully.

{\bf III.}\,\,Let us remind that within SVZ \cite{svz} approach $\Pi_{V-A}$
has the theoretical expression
\begin{equation}\label{ope}
\Pi_{V-A}(-Q^2)=-\frac{W_1}{Q^2}+\frac{W_2}{Q^4}+\frac{C^6<O^6>}{Q^6}+
\frac{C^8<O^8>}{Q^8}+...
\end {equation}
where by $W_1,W_2$ we denote dimension 2,4 operators which vanish in the
chiral limit, $C^6<O^6>$ comes from the four-quark operators
and $C^8<O^8>$ is the contribution of
dimension $D=8$ operators. Note that pure perturbative contributions
cancel in the difference of vector and axial-vector current correlators
(\ref{ope}).

Using the dispersion representation for $\Pi_{V-A}(Q^2)$ it is easy to
obtain a set of finite energy sum rules:
\begin{equation}\label{w1}
W_1=\int_0^{s_0}\rho(s)ds
\end {equation}
\begin{equation}\label{w2}
W_1=\int_0^{s_0} s\rho(s)ds
\end {equation}
\begin{equation}\label{c6}
C^6<O^6>=-\int_0^{s_0} s^2\rho(s)ds
\end {equation}
\begin{equation}\label{c8}
C^8<O^8>=\int_0^{s_0} s^3\rho(s)ds
\end {equation}
where $s_0$ is the onset of asymptotic regime. Since asymptotically
$\rho(s)\rightarrow 0$, $s_0$ in eqs.(9-12) can be replaced by infinity.
Then the first two of these equations coinside with the two Weinberg
sum rules \cite{wsr}
while the last two are the FESRs used for determining corresponding
condensates \cite{wef}.

One can easily notice that (\ref{r}) is essentially the combination of FESRs
(\ref{w1}),(\ref{c6}) and (\ref{c8}) with the upper limit
of integration substituted
by $M_{\tau}^2$. This substitution however is numerically unimportant due to
the vanishing of $\rho(s)$ at high $s$ values and the double zero
of the integration weight at $s=M_{\tau}^2$. With account of (\ref{w1}),
(\ref{c6}),(\ref{c8}) equation (\ref{r}) can be rewritten as
\begin{equation}\label{rf}
R_{\tau,V-A}=12\pi^2({\frac{W_1^F}{M_\tau^2}}
+\frac{3C^6<O^6>^F}{M_{\tau}^6}+\frac{2C^8<O^8>^F}{M_{\tau}^8})
\end {equation}
where we have used the superscript {\small F} to indicate that the
corresponding quantities are determined through FESRs. Let us note that
dimension $D=8$ contribution, neglected in \cite{bbb}, is still important
in our case.

In  \cite{bbb} the authors essentially calculate the instanton
contributions to FESRs like (\ref{w1}-\ref{c8}) in the vector and
axial-vector channels and their combination (\ref{r}). Their results imply
that equations  (\ref{w1}-\ref{c8}) are strongly modified by
contributions from dimension 18 term in (\ref{ope}) due to instanton effects.
Comparison of equations (\ref{dour}-\ref{dbbb}) could lead to the conclusion
that the standard condensate contributions to FESRs (\ref{w1}-\ref{c8})
are much less than that of small size instantons. In this case the condensates
should be
far smaller than their presently accepted values, obtained mostly by FESRs.
However this would cause a sharp disagreement with the Borel transformed SR
where the corresponding contributions are factorially suppressed. Now we
are going to check this effect.

Let us consider the BTSR for $\Pi_{V-A}$
\begin{equation}\label{b1}
\frac{W_1}{M^2}-\frac{W_2}{M^4}-\frac{C^6<O^6>}{2M^6}
-\frac{C^8<O^8>}{6M^8}+\dots=
\int \rho(s)\exp(\frac{-s}{M^2}){ds\over{M^2}}
\end {equation}
and for $Q^2\Pi(Q^2)$
\begin{equation}\label{b2}
\frac{W_2}{M^4}+\frac{C^6<O^6>}{M^6}
+\frac{C^8<O^8>}{2M^8}+\dots=
\int \rho(s)\exp(\frac{-s}{M^2}){s\over{M^2}}{ds\over{M^2}}
\end {equation}
where the dots stand for contributions of higher dimension operators
which are suppressed factorially.

Neglecting the contributions of dimension $D\geq 10$ and
taking different values of the Borel parameter $M^2$ in equations (\ref{b1}),
(\ref{b2}) we are able to construct exactly the same combination of
condensates as the one entering eq.(\ref{rf}):
\begin{eqnarray}\label{rb}
R_{\tau,V-A}^B & \equiv & 12\pi^2({\frac{W_1^B}{M_\tau^2}}
+\frac{3C^6<O^6>^B}{M_{\tau}^6}+\frac{2C^8<O^8>^B}{M_{\tau}^8})  \nonumber\\
               & =      &
\frac{12\pi^2}{M^2}\int [\exp({{-s}\over{M_1^2}})+
{{s}\over{M_1^2}}\exp({{-s}\over{M_2^2}})]\rho(s) \,ds
\end{eqnarray}
where $M_1^2=M_{\tau}^2/5.18 \approx 0.6GeV^2$, $M_2^2=M_{\tau}^2/3.17
\approx 1GeV^2$.

According to ref.\cite{bbb} the instanton contributions in (\ref{rb})
should be negligible, being suppressed by $7!$. Hence
the difference of eqs.(\ref{r}) and (\ref{rb}) is just the measure of
instanton contributions calculated in \cite{bbb}
\begin{equation}
R_{\tau,V-A}^{inst}  =  R_{\tau,V-A}-R_{\tau,V-A}^B
\end{equation}

Taking the difference of  (\ref{r}) and (\ref{rb}) and performing
integration with the measured spectral density \cite{web,wef} we can now
evaluate

\begin{eqnarray}\label{delt}
R_{\tau,V-A}^{inst} & = & {{12\pi^2}\over{M_{\tau}^2}}
\int [1-{{3s^2}\over{M_{\tau}^4}}+{{2s^3}\over{M_{\tau}^6}}-
\exp({{-s}\over{M_1^2}})-{{s}\over{M_1^2}}\exp({{-s}\over{M_2^2}})]\rho(s)
\,ds \nonumber   \\    &= & 0.018 \pm 0.014
\end{eqnarray}

This result can be improved even further if we use the BTSR for
$Q^4\Pi(Q^2)$ as well. In this case we can construct the representation of
$R_{\tau,V-A}^B$ with zero coefficient of the dimension $D=10$ condensate,
which is presumably the dominant correction neglected in (\ref{delt}).
In this case the instanton corrections should be still suppressed by 6!.
The result for $R_{\tau,V-A}^{inst}$, analogous to (\ref{delt}),
 looks like:
\begin{eqnarray}\label{rins}
R_{\tau,V-A}^{inst} &= & {{12\pi^2}\over{M_{\tau}^2}}
\int [1-{{3s^2}\over{M_{\tau}^4}}+{{2s^3}\over{M_{\tau}^6}}-
\exp({{-s}\over{M_1^2}})-{{s}\over{M_1^2}}\exp({{-s}\over{M_2^2}})+
\beta {{s^2}\over{M_3^4}}\exp({{-s}\over{M_3^2}})]\rho(s)\,ds  \nonumber \\
       &= & (1 \pm 3)\cdot 10^{-3}
\end{eqnarray}
where $M_1^2=M_{\tau}^2/2.95 \approx 1.1GeV^2$, $M_2^2=M_{\tau}^2/1.5
\approx 2.1GeV^2$, $M_3^2=M_{\tau}^2/1.01 \approx 3GeV^2$ and $\beta=2.86$.
Note that in this equation D=12 term is also suppressed, stronger than say
in eq.(\ref{b1}.

Taking three standard deviations in (\ref{rins}) we conclude that
\begin{equation}
R_{\tau,V-A}^{inst}<0.01
\end{equation}
which is at least an order of magnitude less than (\ref{dbbb})

Assuming that the relation obtained in \cite{bbb}
\begin{equation}
R_{\tau}^{inst} \equiv R_{\tau,V+A}^{inst} \simeq
{\frac{1}{20}}R_{\tau,V-A}^{inst}
\end{equation}
is still valid, we obtain the following estimate for instanton contributions
to the $\tau$ decay width
\begin{equation}
R_{\tau}^{inst}<5 \cdot 10^{-4}
\end{equation}
which is at least an order of magnitude less than the result of \cite{bbb}
and even much less than standard condensate contributions evaluated in
\cite{bnp}.

{\bf IV.}\,\,\, Thus, analysing the sum rules for $\Pi_{V-A}$, at the
present level of accuracy we can see no sign of discrepancy between
FESR and BTSR which could signal the presence of instanton contributions.
Note that such contributions could lead to serious inconsistencies
in BTSR for $\pi \rightarrow e \nu\gamma$ decay axial form factor
and for the pion electromagnetic mass difference \cite{I} which are also
determined by $\Pi_{V-A}$.

Instanton contributions to the $\tau$ hadronic width are smaller than
the standard condensate corrections by more than order
of magnitude and thus arise no difficulties for improvement of accuracy
in $\alpha_s$ determination.
The main problem still is the evaluation of $O(\alpha_s^4)$
perturbative contributions \cite{kat}.

{\bf V.}\,\,\, We have benefited from helpful discussions with
Z.Berezhiani.

One of us (M.M.) wants to express his gratitude to
Z.Berezhiani and G.Fiorentini for their warm hospitality at Ferrara
section of INFN where part of this work was done.

The work of M.Margvelashvili was in part supported by the grant of
International Science Foundation.


\begin{thebibliography}{99}
\bibitem{bnp}
E.Braaten, S.Narison and A.Pich, Nucl.Phys. B 373 (1992) 581.
\bibitem{dib}
F.Le Diberder and A.Pich, Phys.Lett. B 289 (1992) 165
\bibitem{p}
A.Pich CERN preprint CERN-TH.6738/92 (1992).
\bibitem{al}
ALEPH Collab. Busculic et al., Phys.Lett. B 307 209 (1993).
\bibitem{svz}
M.A.Shifman, A.I.Vainshtein and V.I.Zakharov, Nucl.Phys. B147, 385 (1979).
\bibitem{np}
P.Nason and M.Poratti
CERN preprint CERN-TH. 6738/92 (1992).
\bibitem{bbb}
I.I.Balitsky, M.Beneke and V.M.Braun, Phys.Lett. B 318, 371 (1993)
\bibitem{fesr}
K.G.Chetyrkin, N.V.Krasnikov and A.N.Tavkhelidze, Phys.Lett. 76B 83 (1978),
K.G.Chetyrkin and N.V.Krasnikov Nucl.Phys B119, 174 (1977).
\bibitem{web}
V.G.Kartvelishvili and M.V.Margvelashvili, Yad.Fiz. 52 (1990).\\
\bibitem{wef}
V.G.Kartvelishvili and M.V.Margvelashvili, Z.Phys. C 55, 83 (1992).
\bibitem{wsr}
S.Weinberg, Phys.Rev.Lett. 18 507 (1967)
\bibitem{I}
M.Margvelashvili, Phys.Lett B 215 763 (1988);
M.Margvelashvili, Int.J.Mod.Phys. A 5 747 (1990).
\bibitem{kat}
A.L.Kataev and V.V.Starshenko Preprint CERN-TH.7198/94 (1994)
\end{thebibliography}
\end{document}